# Manipulate intrinsic light-matter interaction with bound state in the continuum in van der Waals metasurfaces by artificial etching

Fuhuan Shen[1], Xinyi Zhao[1], Yungui Ma[2], Jianbin Xu[1,3*]

1. Department of Electronic Engineering, and Materials Science and Technology Research Center, The Chinese University of Hong Kong, Shatin, N.T., Hong Kong SAR, P. R. China.
2. State Key Lab of Modern Optical Instrumentation, Centre for Optical and Electromagnetic Research, College of Optical Science and Engineering; International Research Center (Haining) for Advanced Photonics, Zhejiang University, Hangzhou 310058, China
3. State Key Laboratory of Quantum Information Technologies and Materials, The Chinese University of Hong Kong, Shatin, N.T., Hong Kong SAR, P. R. China.

* Corresponding Author. Email: jbxu@ee.cuhk.edu.hk (J. B. Xu)

**Abstract**

The recent demonstrations of van der Waals (vdW) nanophotonics have opened new pathways for manipulating the light-matter interaction in an intrinsic manner, leading to fascinating achievements in tunable magneto-optics by self-hybrid polaritons, indirect bandgap lasering, and exceptionally enhanced optical nonlinearity. However, the anisotropic atomic lattice, chemically active side walls, and distinct enthalpies of formation across vdW materials, pose significant challenges in nanofabrication and material choices, hindering the realization of high-$Q$ resonant mode on arbitrary materials. In this work, we propose an etch-free vdW structure that mimics the shallow etching, termed "artificial etching." This approach utilizes a low refractive index (LRI) perturbation layer made of photoresist, drastically reducing radiation loss and experimentally achieving a remarkable $Q$ factor of up to 348—comparable to the highest values reported in vdW nanophotonics. We demonstrate room-temperature polaritons in etch-free structures using four representative materials ($WS_2$, $MoS_2$, $WSe_2$, and $MoSe_2$) through self-hybridization of high-$Q$ (quasi-)bound states in the continuum (BIC) modes and excitons, achieving a Rabi-splitting of approximately 80 meV, which significantly surpasses the intrinsic excitonic loss. Furthermore, we showcase optical modulation of indirect bandgap emission in bulk $WS_2$ and direct exciton emission in heterostructures, achieving substantial polarization-dependent enhancement of their emission efficiencies. The proposed etch-free vdW structure provides a versatile platform for high-$Q$ nanophononics while preserving material integrity, advancing applications in photoelectronic and quantum devices.

**Introduction**

Modern photoelectronic and quantum applications — such as nanolasers, photodetectors, and ultrafast optical switches—heavily depend on the precise manipulation and enhancement of light-matter interactions[1-5]. However, the interaction between natural materials and light is inherently constrained by their atomic composition and structural arrangement, which often limits their performance in practical applications. A promising solution lies in integrating materials with photonic nanostructures that support highly resonant modes, enabling spatial and spectral overlap to tailor and amplify these interactions[6,7,8]. Among the various photonic modes, the bound state in the continuum (BIC) has garnered significant attention due to its unique properties. BICs describe nonradiative photonic states that exist within the radiation continuum (inside the light cone), offering exceptional quality ($Q$) factors, tightly confined near fields, and rich topological characteristics[9-12]. To date, experimental demonstrations have successfully hybridized BIC modes in photonic structures with materials such as dye molecules[11,13], showing promising influences in areas such as biodetection[14,15], enhanced light harvesting and emission[11,13], and nonlinear optics[16] for quantum technologies.

The family of two-dimensional (2D) materials—spanning insulating hexagonal boron nitride (hBN), semiconducting transition metal dichalcogenides (TMDCs), and semimetallic graphene—provides a versatile platform for exploring light-matter interactions across a broad spectral range[17]. These materials consist of covalently bonded atomic layers held together by weak van der Waals (vdW) forces, enabling precise control over layer thickness (from monolayers to bulk materials exceeding 100 nm) and offering surfaces free of dangling bonds. This unique structural flexibility makes 2D materials ideal for integration with a wide range of photonic structures, such as waveguides[18] and photonic crystals[19,20], and allows them to be transferred onto virtually any flat substrate. Moreover, the inherent 2D nature of vdW materials grants them extraordinary photonic and electronic properties that are unattainable in their 3D counterparts, such as silicon. For example, monolayer TMDCs exhibit room-temperature excitons with direct bandgap emission and strong absorption rates of approximately 10%[21,22], while hBN and graphene support intrinsic polaritons enabled by phonons and plasmons[23]. These distinctive features significantly expand the toolkit for manipulating light-matter interactions, paving the way for groundbreaking scientific discoveries and advanced applications.

The emerging field of hybridizing BIC modes in dielectric photonic structures with vdW materials has recently demonstrated remarkable achievements, including enhanced linear and nonlinear light sources[16,24], vortex beam generation[25], and strongly coupled exciton-polaritons[26-29]. Maximizing these interactions requires both strong field confinement via the BIC mode and placement of the material at the field maximum[30]. However, in many hybrid systems, 2D materials are positioned adjacent to the external photonic structure, often separated by a buffer layer of hexagonal boron nitride (hBN)[16], interacting to the evanescent field near the surface.

Currently, many previously reported vdW photonic structures exhibit relatively low-quality (Q) modes due to significant radiation losses and fabrication imperfections[31-36]. Recently, the concept of (quasi-)bound states in the continuum (BIC) has been applied to vdW metasurfaces, demonstrating controlled $Q$ factors ($Q > 300$) and enhanced light-matter interactions by precisely tuning the asymmetric parameter and scaling factor of the structure[37,38]. Nevertheless, achieving this level of control demands exceptionally high fabrication precision. The anisotropic lattice symmetry and varying enthalpies of formation across different materials pose significant fabrication challenges, often resulting in uncontrolled shapes and low precision in the final nanostructures after standard etching processes[39]. These issues are exacerbated for smaller dimensions and thicker materials[36]. Additionally, the chemically active sidewalls created by etching are more reactive than the pristine vdW surfaces, leading to undesirable chemical reactions in materials such as $MoS_2$ and a reduction in excitonic oscillator strength[36,40]. Together, these factors impose severe limitations on vdW BIC metasurfaces, restricting material choices (to date, only hBN and $WS_2$ metasurfaces have demonstrated high-Q BIC modes[37,38]) and requiring stringent fabrication conditions. These constraints hinder broader applications, such as in nonlinear quantum sources and anisotropic photonic devices where many other vdW materials will be involved.

In this work, we propose an alternative to traditional "real etching" by introducing an etch-free vdW metasurface that achieves intrinsic light-matter interaction through a perturbed layer made of a low-refractive-index (LRI) photoresist. This approach mimics the optical behavior of shallow-etched vdW metasurfaces, a concept we term "artificial etching." As a proof of concept, we implement a grating structure that supports guided mode resonance (GMR) and BIC modes, achieving simulated $Q$ factors exceeding $10^3$ and measured $Q$ factors (at normal incidence) approaching 348—comparable to the highest values reported in etched BIC vdW metasurfaces[37,38]. Using this etch-free platform, we demonstrate self-hybridized exciton-polaritons in four representative TMDCs ($WS_2$, $MoS_2$, $WSe_2$, and $MoSe_2$) through angle-resolved transmission measurements, with results that align perfectly with simulations. Furthermore, the high-$Q$ modes enable over 25 times enhancement and narrowed linewidth of ~13 nm in indirect bandgap emission from bulk $WS_2$, with a calculated Purcell factor exceeding 200. Additionally, we apply the etch-free vdW structure to a heterostructure composed of a monolayer (ML) $MoSe_2$ encapsulated in hBN layers, achieving plexcitons through weak coupling and demonstrating significant modulation of the emission polarization of direct excitons in ML $MoSe_2$. The proposed etch-free vdW structure is universally applicable to arbitrary vdW materials and their homo-/heterostructures, unlocking new possibilities in vdW nanophotonics and light-matter interactions. This approach paves the way for novel devices in nonlinear optics, lasing, and quantum information technologies.

**Results**

**Concept and basic optical properties**
As schematically illustrated in Fig. **1a**, the etch-free structure consists of a top layer made of a low-refractive-index (LRI) photoresist (e.g., ZEP 520A in this work) and a bottom van der Waals (vdW) layer, such as bulk $WS_2$. The periodic structure formed by the LRI polymer perturbates the in-plane continuous translation symmetry along the vdW surface, converting the guided mode within the dielectric layer (i.e., the vdW layer) into a leaky mode, known as guided mode resonance (GMR). The leaky rate, or scattering radiation, is determined by the perturbation introduced by the periodic structure. Supplementary Fig. **S2** demonstrates the evolution of the scattering rate (inverse of the $Q$ factor for non-loss dielectric materials) with respect to the etching depth of a general dielectric grating. As the etching depth increases, the scattering radiation is significantly enhanced, leading to a pronounced reduction in the $Q$ factor. In this context, the introduction of the LRI nanostructured layer serves a similar function to shallow etching, as it minimally perturbs the photonic properties of the underlying high-refractive-index (HRI) dielectric layer.

To illustrate this, Fig. **1b** compares the transmission spectra of the etch-free structure and a shallow-etched structure both made from 60 nm WS$_2$. The polymer thickness in the etch-free structure is set to 50 nm, while the dielectric structure is shallowly etched by 3 nm for comparison. Both structures share identical parameters, including period ($a$ =480 nm) and filling factor ($w = a$ /2). The transmission spectra for TE and TM modes exhibit nearly identical responses in terms of resonant frequencies, linewidths, and field distributions (see insets in Fig. **1b**). Furthermore, the two structures display similar evolution of optical modes (the TE mode used as a representative example) in the $k_x$ and $k_y$ directions (Supplementary Fig. **S4**), indicating comparable photonic band structures.

The photonic TE band structure of the etch-free structure is depicted in the $k_x$-$k_y$ space, as shown in Fig. **1c**. Two branches are observed in momentum space, exhibiting near-linear dispersion in the $k_x$ direction while remaining nearly constant in the $k_y$ direction. A zoomed-in view of the band structure in the $k_x$ direction (Fig. **1d**) reveals an anti-crossing behavior between these two branches. According to perturbation theory[41,42], the formation of the guided mode resonance (GMR) and the BIC modes arises from the diffractive coupling of band-folded propagating modes across the Brillouin zone (see Supplementary Fig. **S1** for details). The narrowed gap in Fig. 1d indicates weak coupling strength, which stems from the small perturbation introduced by the LRI nanostructured layer. As a result, in the etch-free structures, the operating frequencies for normal GMR and BIC are almost degenerate and become separate at inclined incidence.

The corresponding near-field distributions (bottom of Fig. **1d**) illustrate the distinct photonic behaviors of these modes. The BIC mode exhibits an asymmetric near-field distribution, leading to total destructive interference in the far field and rendering it optically dark under normal incidence. In contrast, the GMR mode displays a fully symmetric distribution. As shown in Fig. **1e**, the $Q$ factor for the BIC branch decreases dramatically from an infinite value to a finite one as $k_x$ increases, while the GMR mode exhibits the opposite trend. Both modes would share almost the same linewidth ($\gamma_r$) at finite (but very small) $k_x$. (see details in Supplementary Note **S1**). These two branches are well reproduced in the angle-resolved transmission spectra, both experimentally and through simulations, for a representative WS$_2$ etch-free structure, as shown in Fig. 1f. In contrast to the BIC and GMR modes, the transmission dip associated with A exciton absorption (indicated by the green dashed line) shows no dispersion with $k_x$. The TM photonic band structure, presented in Supplementary Fig. **S5**, exhibits a similar dispersive behavior and $Q$ factor distribution to the TE mode. Throughout this work, TE and TM modes are selectively analyzed based on their operating frequencies.

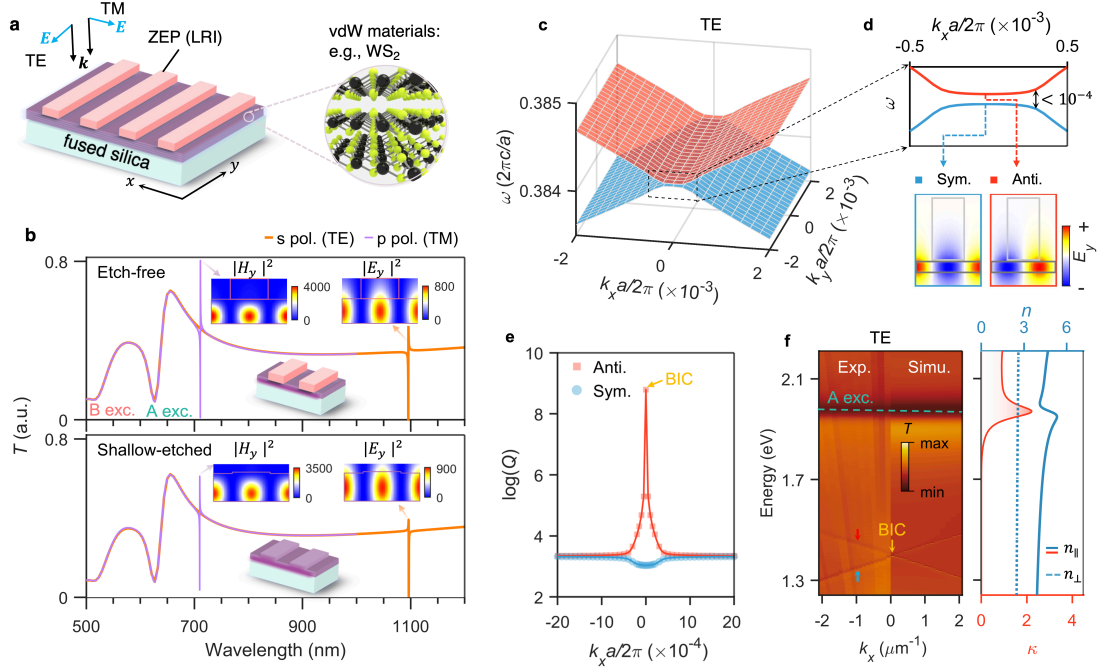

Figure 1. **Proposed etch-free structure and its optical properties.** (**a**) Schematic illustration of etch-free structure based on vdW materials. (**b**) Comparison of TE (orange lines) and TM (purple lines) modes from etch-free and shallow-etched vdW gratings with their geometric parameters are the same except the thickness of the polymer (top) and etched vdW layer (bottom). The insets are the corresponding near-field distributions for these modes. (**c**) The photonic band structure of TE mode in $k_x$-$k_y$ space. (**d**) Zoom-in band structure in $k_x$ (top) direction and corresponding eigen modes (bottom). (**e**) $Q$ factor distributions for the eigen modes shown in (**d**). (**f**) Experimentally measured and simulated angle-resolved transmission spectra (left) of $WS_2$ etch-free structure where the period $a$=350 nm and filling factor $\Lambda$=0.5. The right image shows the in-plane and out-of-plane refractive index of bulk $WS_2$.

**Experimental realization and geometric parameters analyses**

To fabricate etch-free vdW metasurfaces, bulk TMDC or hBN is first exfoliated using scotch tape and transferred to a marked area indicated by Au arrows on a $SiO_2$ (300 nm)/Si substrate. The thickness of the exfoliated material is confirmed through atomic force microscopy (AFM) measurements. The sample is then spin-coated with a layer of ZEP 520A photoresist at 3500 rpm. Subsequently, standard electron beam lithography (EBL) is performed, followed by a developing process to create the final nanopattern on the target vdW layer (see Supplementary Fig. **S11** for more details). This process results in a final ZEP pattern thickness of approximately 300 nm. Fig. **2a** shows a representative sample, with square areas indicating the grating patterns. Several critical geometric parameters—such as the thickness of the ZEP layer ($t_1$), the TMDC layer ($t_2$), the period ($a$), and the width ($w$) or filling factor ($\Lambda$) of the grating—significantly influence the optical performance of the metasurfaces. A comprehensive analysis of these parameters is presented in the following sections. Without loss of generality, bulk $WS_2$ is used as a representative example throughout this study.

The period of the photonic structure dictates the resonant wavelength of the mode. As the period increases from 343 nm to 503 nm (shown in scanning electronic microscope (SEM) images,

Fig. **2b**), the optical mode exhibits a pronounced, near-linear redshift in both measured (Fig. **2c**) and simulated (Fig. **2d**) results. This confirms that the resonant modes originate from the periodic ZEP structure rather than trivial effects.

Another key consideration is the critical thickness of the vdW layer required to support a pronounced resonant mode in the etch-free structure. Fig. **2e** shows the evolution of transmission spectra with WS$_2$ thickness for a period $a$=450 nm and filling factor $\Lambda$=0.5. For WS$_2$ thicknesses below 20 nm, only negligible optical responses are observed, attributed to Rayleigh anomalies (RAs), which are frequency-independent and determined by the period and environmental refractive index[43]. However, as the thickness increases beyond ~35 nm, a distinct GMR mode emerges, exhibiting a clear redshift with further thickness increases. To experimentally validate the influence of TMDC thickness, a metasurface with uniform geometric parameters ($a$=450 nm, $\Lambda$=0.5) was fabricated on WS$_2$ layer region (Fig. **2f**) of two thicknesses (~60 nm and ~70 nm, confirmed by AFM, Fig. **2j**). The measured transmission spectra (Fig. **2i**) align perfectly with simulations (Fig. **2h**) extracted from Fig. **2e**, confirming that the thickness exceeds the critical value and demonstrating the thickness ($t_2$)-dependent behavior of the GMR mode.

The thickness of the ZEP layer ($t_1$) also influences the optical response. While the experimental samples maintain a uniform ZEP thickness of 300 nm, a detailed analysis of $t_1$ via simulation is provided in Supplementary Fig. **S6**. Intuitively, reducing the ZEP thickness increases the Q factor due to lower scattering radiation. Recent study[44] using a 50 nm PMMA etch-free structure have achieved $Q$ factors exceeding $10^6$. However, thinner LRI layers pose experimental challenges due to reduced mode amplitude[45] (thus the increase of signal-to-noise ratio), creating a trade-off between $Q$ factor and measurement intensity when selecting the optimal LRI thickness. Additionally, we observe that for ZEP thicknesses above 150 nm, the optical mode remains largely unchanged. Further optimization of the photoresist thickness could be made to enhance the $Q$ factor.

In addition to ZEP thickness, the filling factor ($\Lambda$) also determines scattering radiation. As shown in Fig. **2j**, $\Lambda$=0.5 corresponds to the highest radiation loss (maximum perturbation). Deviating from this value, either increasing or decreasing $\Lambda$, unambiguously increases the $Q$ factor of the mode, as confirmed by simulated transmission spectra (Fig. **2k**). Experimentally measured transmission spectra (Fig. **2l**) align with this trend, showing nearly unchanged resonant frequencies but reduced linewidths when $\Lambda$ deviates from half the period.

Thanks to the low radiation loss in the etch-free structure, the $Q$ factors for our vdW metasurfaces reach ~60-100 for TM modes (Fig. **2i**) and several hundred for TE modes (Fig. **2l**, with $Q$ = 348 in Fig. **4b**). These values are comparable to the highest $Q$ factor (~370) achieved in recent etched symmetry-protected BIC metasurfaces[38]. Deviations between measured and simulated results primarily arise from (i) fabrication imprecision and (ii) the use of an incoherent light source. Previous studies on plasmonic surfaces suggest that coherent source excitation could double the $Q$ factor compared to incoherent excitation[46].

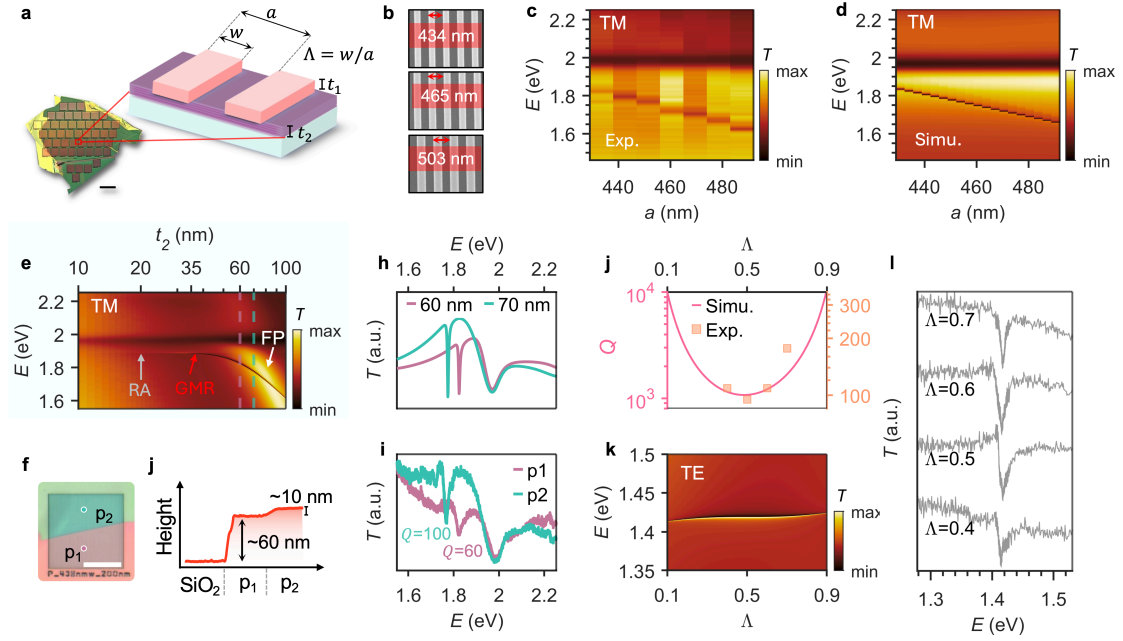

Figure 2. The dependence of geometric parameters on optical properties. (**a**) Microscopic image of one representative etch-free TMDC metasurface and its corresponding schematical structures. The scale bar is 50 μm. (**b**) SEM images showing different grating periods of ZEP nanostructures. (**c-d**) Measured (**c**) and simulated (**d**) transmission spectra for etch-free metasurfaces with varied periods under TM polarization. Other parameters including filling factor are the same. (**e**) Evolution of optical mode with TMDC thickness $t_2$. (**f**) Microscopic image of structure with thickness WS$_2$ of varied. (**j**) Measured height for different thicknesses in (**f**) by AFM. (**h-i**) Simulated and measured transmission spectra for WS$_2$ of different thicknesses. (**j**) Evolution of $Q$ factor with the filling factor $\Lambda$. (**k**) Evolution of transmission spectra with filling factor $\Lambda$. (**l**) Measured transmission spectra for $\Lambda$ from 0.4 to 0.7.

**Self-hybridized polaritons in various TMDC materials**

According to the above analyses, we prepare four types of TMDC materials, namely, WS$_2$, MoS$_2$, WSe$_2$, and MoSe$_2$, with their thicknesses ranging from ~60 nm to ~80nm (Supplementary Fig. **S12**). These vdW materials all possess high refractive index with their excitonic response varied from visible (WS$_2$: ~620 nm) to the near-infrared range (MoSe$_2$: ~800 nm) as indicated in the right panels of Fig. **3a-d**. Through choosing the suitable periods for the corresponding materials, the hybridization of photonic modes, i.e., GMR and BIC, with their *A* excitons (due to far detuning the coupling with B excitons is neglected for the sake of simplicity) have been observed in both measured and simulated angle-resolved transmission spectra (Fig. **3a-d**).

The strong couplings between photonic modes with the *A* exciton can be described by the eigen function:

$$H_{SC}|\varphi\rangle = E|\varphi\rangle, \quad (1)$$

where $|\varphi\rangle$ is eigen state of polariton state and $E$ is the corresponding eigen energy. $H_{SC}$ describes the strong coupling Hamiltonian for the hybrid system whose explicit form can be expressed as

$$H_{SC} = \begin{bmatrix} \Omega_{BIC} & g & 0 & 0 \\ g & \Omega_X & 0 & 0 \\ 0 & 0 & \Omega_{GMR} & g \\ 0 & 0 & g & \Omega_X \end{bmatrix}, \quad (2)$$

where $\Omega_{(BIC,BMR)} = \omega_{(BIC,GMR)} + i\gamma_{(BIC,GMR)}$ is the eigen frequency of BIC or GMR mode and $g$ is the coupling strength. $\Omega_X = \omega_x + i\gamma_x$ is the eigen frequency of A exciton. The total Hamiltonian can be decomposed into two 2 × 2 block matrixes, each representing the interaction of the $A$ exciton with the BIC and GMR modes, respectively.

The eigen state is the hybrid state composed of photonic and excitonic constituents, i.e., $|\varphi\rangle = \alpha|\varphi_{Pho.}\rangle + \beta|\varphi_X\rangle$ where $\alpha$ and $\beta$ are the Hopfield coefficients. Fig. **3e-h** exhibits the corresponding polaritonic energy dispersion to Fig. **3a-d**. The BIC mode is spectrally overlapped with exciton with $k_x$, showing pronounced anti-crossing behavior and evolution of Hopfield coefficients (i.e., change of photonic and excitonic portions in polariton states). Due to the far detuning, GMR modes shows only negligible hybridization with A excitons. By comparing Fig. **3a-d** with Fig. **3e-h**, we found that although the dispersions of LP$_{BIC}$ branches are pronounced in all transmission spectra in Fig. **3a-d** while UP$_{BIC}$ branches are only observed in the WS$_2$ and MoSe$_2$ samples in both measured and simulated results. This could be possibly due to the large absorption by $B$ exciton (MoS$_2$) and relatively small oscillator strength of $A$ exciton (MoSe$_2$)[28,30]. As a result, the Rabi splitting, simply defined as $\Omega = 2g$, is extracted as 80 meV for WS$_2$ and as 72 meV for MoSe$_2$, significantly exceeding the linewidths of their respective $A$ excitons ($\gamma_{WS_2}$=35 meV and $\gamma_{MoSe_2}$=43meV).

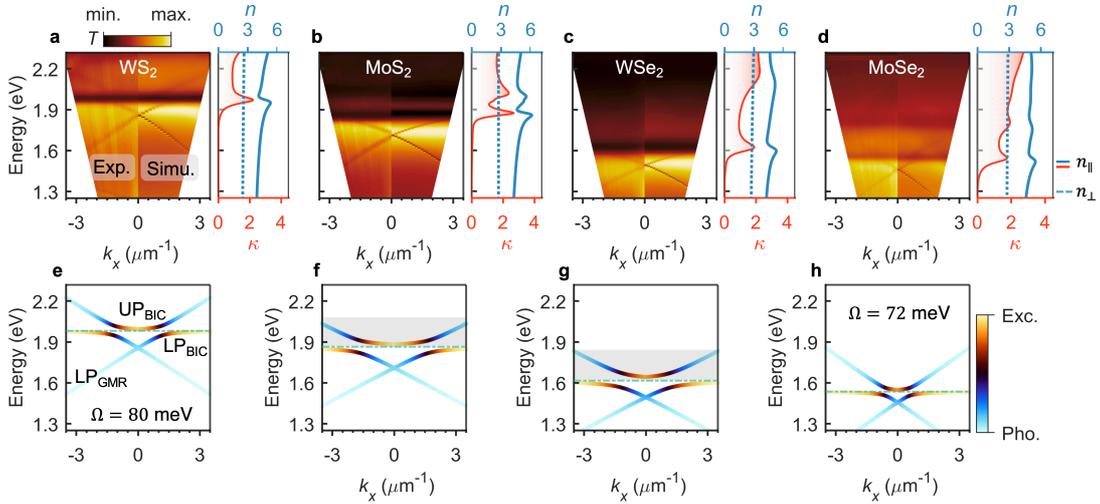

Figure 3. Self-hybrid polaritons in four TMDCs. (**a-d**) Measured and simulated angle-resolved transmission spectra for etch-free WS$_2$, MoS$_2$, WSe$_2$, and MoSe$_2$ metasurfaces. Filling factors for these metasurfaces are fixed at 0.5. The right panels show their corresponding refractive index. (**e-h**) Corresponding calculated polariton dispersions with the colormap indicating their photonic and excitonic partition in the hybrid polariton states. The UP$_{GMR}$ mode is removed for clarity of image display which will not affect the analyses in the main text. All the results are from the TM modes.

**Enhancement of the indirect bandgap emission**
Most bulk TMDCs exhibit indirect bandgaps due to momentum mismatch between the valence band maximum and conduction band minimum, as illustrated in Fig. **4c**. This results in direct and indirect bandgap emissions, leading to relatively low quantum efficiency (optical dark mode) compared to their monolayer counterparts. Consequently, prior research on TMDC photonic structures has primarily focused on passive devices. However, the strong field enhancement and confinement provided by GMR and BIC modes offer the potential to brighten these dark modes, enabling novel applications such as lasing, as demonstrated in WS$_2$ disk-based whispering gallery cavities[47].

Figures **4a-b** shows a high-Q TE mode at around 890 nm in an etch-free WS$_2$ metasurface, spectrally overlapping with the indirect bandgap emission (Fig. **4d**). As shown in Fig. **4e**, the indirect emission is enhanced by over ~25 times, with the linewidth narrowed to ~13 nm due to the large Purcell effect confined within the WS$_2$ layer (inset of Fig. **4e**). Additionally, the emission from the etch-free nanostructure exhibits strong linear polarization, with $I_{parallel}/I_{cross}$=12.5, where $I_{parallel}$ and $I_{cross}$ represent emission intensities with the electric field aligned parallel and perpendicular to the grating bars, respectively (as indicates in Fig. **5a**).

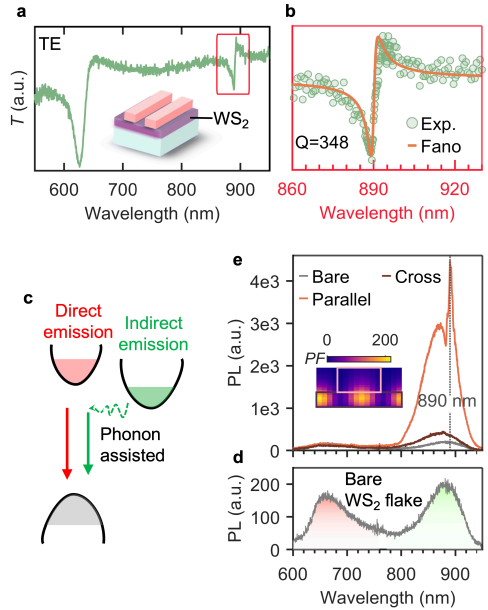

Figure 4. Enhancement of indirect-bandgap emission by etch-free structure. (**a**) Transmission spectra for etch-free WS$_2$ metasurface. (**b**) Zoom-in spectrum (green circles) in (**a**) showing high-Q mode by the Fano fitting (orange line). (**c**) Schematic illustration of direct and indirect emission in bulk WS$_2$. (**d**) The PL spectrum for bare WS$_2$ flake where the red and green areas indicate the direct and indirect emissions. (**e**) Enhanced PL by the etch-free structure. The emissions with electric field parallel (orange line) and cross (brown line) to the grating bar direction are separated.

**Intrinsic interaction in heterostructure**

Maximizing the interaction between photonic modes in dielectric cavities and direct excitons in monolayer (ML) TMDCs or Moiré excitons in TMDC heterostructures has profound implications for quantum and photoelectronic applications. Previous simulations have shown that placing ML TMDCs inside dielectric cavities significantly enhances light-matter interaction[30], though experimental realization remains challenging[48]. Here, we demonstrate an etch-free structure based on a heterostructure, as illustrated in Fig. 5a. The heterostructure consists of a ML MoSe$_2$ layer sandwiched between top and bottom hBN layers.

Figure 5c shows the measured transmission spectra for five metasurfaces (Fig. 5b) and the photoluminescence (PL) spectrum (red) from ML MoSe$_2$ (without metasurfaces). Compared to TMDC-based structures, the etch-free heterostructure exhibits a reduced $Q$ factor, attributed to the rough surface of the heterostructure (lower fabrication precision) and the relatively low refractive index of hBN (increase of the leaky to the environment). These limitations can be addressed by optimizing heterostructure preparation, reducing photoresist thickness, and adjusting the filling factor, as highlighted in Fig. **2**.

As shown in Fig. 5c, the spectrum for $a$=470 nm broadens when it spectrally overlaps with the $A$ excitonic transition of MoSe$_2$. Meanwhile, a dip (marked by red arrows) appears in other spectra for different periods, indicating weak coupling between light and matter. In this regime, spontaneous emission is accelerated by the enhanced local density of states due to photonic resonant modes, a phenomenon known as the Purcell effect. As shown in the PL mapping (Fig. **5d**), the PL emission over the metasurfaces is enhanced by approximately 3–5 times, consistent with the Purcell factor formula $\frac{\gamma_c}{(\omega_X-\omega_c)^2+\gamma_c}$ where $\omega_{(c,X)}$ is the resonant frequency of cavity or exciton and $\gamma_c$ is the linewidth of cavity. In contrast, the cross-polarized PL emission (Fig. 5e) shows a significant reduction due to the polarization behavior of the TE mode. In the future work, the strong coupling regime could be achieved[48] by improving the quality of the TMDC monolayer, optimizing the thickness of the top and bottom hBN layers (to maximize field interaction with the TMDC), and increasing the $Q$ factor of the photonic mode.

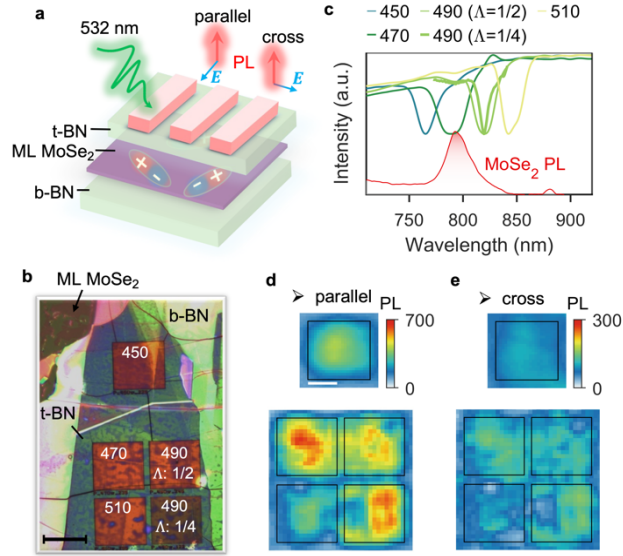

Figure 5. Modulation emission from ML MoSe$_2$. (a) Schematic structure of the heterostructure where a ML MoSe$_2$ layer is sandwiched between a bottom hBN layer (~50 nm) and a top hBN layer (~100 nm). (b) Microscopic image of etch-free heterostructure metasurface. (**c**) Measured transmission for metasurfaces of different periods and filling factors and PL for the uncoupled MoSe$_2$. (d-e) Parallel and cross PL mapping for the metasurfaces.

**Conclusion**

In summary, we propose an etch-free structure for vdW materials, eliminating the need for etching in traditional vdW-based photonic structures. This approach is applicable to arbitrary materials and their heterostructures in principle. High-$Q$ GMR modes are demonstrated both in simulation (>10$^3$) and experimentally (~348) using HRI TMDC materials, achieving $Q$ factors comparable to the record values in vdW nanocavities. Using this structure, self-hybrid polaritons were observed in four TMDC materials, with the largest upper polariton (UP) and lower polariton (LP) energy splitting reaching 80 meV, exceeding the linewidths of both the exciton and BIC modes. Beyond passive devices, we demonstrate modulation of indirect bandgap emission in bulk $WS_2$ and direct bandgap emission in hBN-encapsulated ML $MoSe_2$. The proposed etch-free structure preserves the integrity of vdW materials, enabling the photoelectronic devices with modulated photonic response based on vdW materials.

Recently, the emergent vdW magnets CrSBr[49], 3R $MoS_2$ and $WS_2$ with intrinsic nonlinearity[50,51], and triclinic materials $ReS_2$ and $ReSe_2$[52] present unprecedent opportunities in vdW-based nanophononics for quantum technologies and applications. The propose etch-free vdW structure would expand toolkits in nonlinear optics, photonic spins, and entangled quantum source with intrinsic light-matter interaction

During the submission of this work, we noted a recently published study using an etch-free structure on InSe, achieving remarkable enhancements (>200) in out-of-plane dipole emission and second harmonic generation. While a similar structure was employed, our work distinguishes itself by (i) realizing high-$Q$ GMR and (quasi-)BIC modes, (ii) demonstrating self-hybrid polaritons through intrinsic light-matter interaction across multiple TMDCs (not limited to a single material), and (iii) modulating both indirect bandgap emission in bulk TMDCs and direct bandgap emission in hBN-encapsulated ML TMDCs.

**Methods**

**Simulations**
The normal and angle-resolved transmission spectra are calculated on the home-built Rigorous coupled-wave analysis (RCWA) algorithm. The eigen mode analysis and band structure calculations are solved by the commercial software COMSOL. The Purcell factor is calculated based on the Finite-Difference Time-Domain (FDTD) method by the commercial software Lumerical-FDTD Solutions.

**Fabrications**
The fabrication of etch-free vdW structures on $SiO_2$ (300 nm)/Si is well illustrated in main text and Supplementary **S11**. For the transmission measurements, the vdW structures are wet etched in buffered HF (15%) solvent for 15 mins and then transferred to the transparent substrates. Details can be referred to Ref. 7 in our previous work.

**Optical measurements**
The transmission measurements are based on the combination of the home-built setup and commercial spectrometer (Supplementary Fig. 11). The angle-resolved measurements are realized by the step-motor to precisely control the rotation angle of the sample. The PL measurements are based on the commercial spectroscopic system (HORIBA LabRAM HR Evolution system).

**Acknowledgement**
The work is in part supported by Research Grants Council of Hong Kong, particularly, via Grant

**Supplementary materials** for "Manipulate intrinsic light-matter interaction with bound state in the continuum in van der Waals metasurfaces by artificial etching"

Fuhuan Shen[1], Xinyi Zhao[1], Yungui Ma[2], Jianbin Xu[1,3*]

4. Department of Electronic Engineering, and Materials Science and Technology Research Center, The Chinese University of Hong Kong, Shatin, N.T., Hong Kong SAR, P. R. China.
5. State Key Lab of Modern Optical Instrumentation, Centre for Optical and Electromagnetic Research, College of Optical Science and Engineering; International Research Center (Haining) for Advanced Photonics, Zhejiang University, Hangzhou 310058, China
6. State Key Laboratory of Quantum Information Technologies and Materials, The Chinese University of Hong Kong, Shatin, N.T., Hong Kong SAR, P. R. China.

* Corresponding Author. Email: jbxu@ee.cuhk.edu.hk (J. B. Xu)

# Optical properties of the etch-free vdW nanostructures

**Supplementary Note S1**

According to the perturbation theory[1], the formation of the GMR and BIC modes origins from the diffractive coupling of the band-folded propagating modes (Figure **S1b**) over the Brillouin zone, i.e., $E_\pm = E_0 \pm v_g k_x$ where $k_x = \frac{2\pi}{a}$ and $v_g$ is treated as the constant for the sake of simplicity. Being folded above the light line, each leak to the radiation continuum with a rate $\gamma_r$. Total Hamiltonian to describe their coupling can be given by

$$H_{\text{grating}} = \begin{bmatrix} E_0 & U \\ U & E_0 \end{bmatrix} + i\gamma_r \begin{bmatrix} 1 & \cos\phi \\ \cos\phi & 1 \end{bmatrix}, \qquad (3)$$

where $U$ represents the diffractive coupling and $\cos\phi$ describes the loss exchange in the far field interference. Consequently, the eigen modes is obtained as

$$\Omega_\pm = E_0 + i\gamma_r \pm \sqrt{(v_g k_x)^2 + (U - i\gamma_r \cos\phi)^2}. \qquad (4)$$

At $k_x=0$, the imaginary part of $\Omega_-$ would be zero indicating the non-radiative mode, i.e., BIC. The imaginary part of their combination, i.e., $(\Omega_+ + \Omega_-)=2\gamma_r$ indicates the conversation of total decay rates from BIC and GMR mode. For large $k_x$, GMR and BIC mode would share almost the same linewidth. $\Omega_\pm$ shows a linear dispersion with $k_x^2$ when $k_x \ll U$, and with $k_x$ when $k_x > U$, which is schematically indicated in Figure. **S1c** and Figure **1d** in the main text.

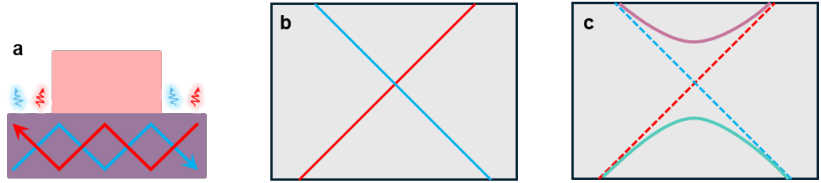

Figure S1. Formation of the BIC and GMR modes in photonic grating. (a) Schematic image of the structure. (b) Folded bands for front and back propagating guided mode. (c) Mode splitting (solid lines) due to the interaction of guided modes (dashed lines) by etched period grating structures.

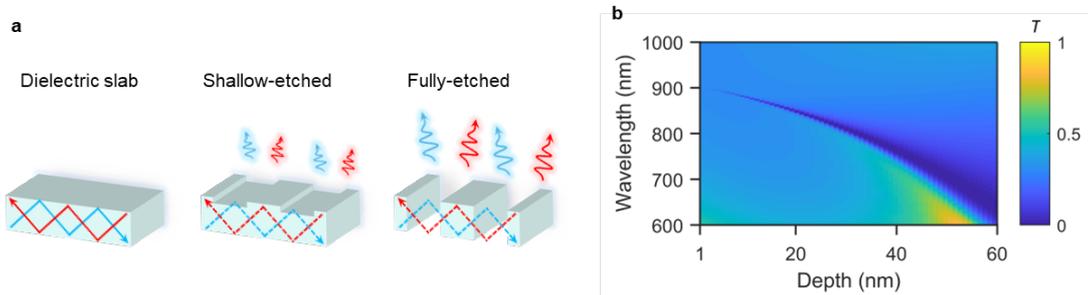

Figure S2. **(a)** Schematic illustration of increased scattering radiation with the etching depth. (b) The evolution of transmission spectra (TE mode) with the etching depth. The period is set as 350 nm, and the original thickness of the dielectric (n=4) layer is set as 60 nm.

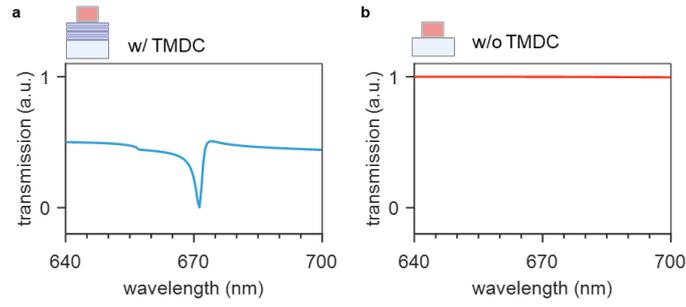

Figure S3. (a-b) Comparison of transmission spectra from the LRI grating structure (pink color) with (**a**) /without (**b**) the HRI TMDC on the bottom. The results indicate the LRI grating structure itself shows no optical response.

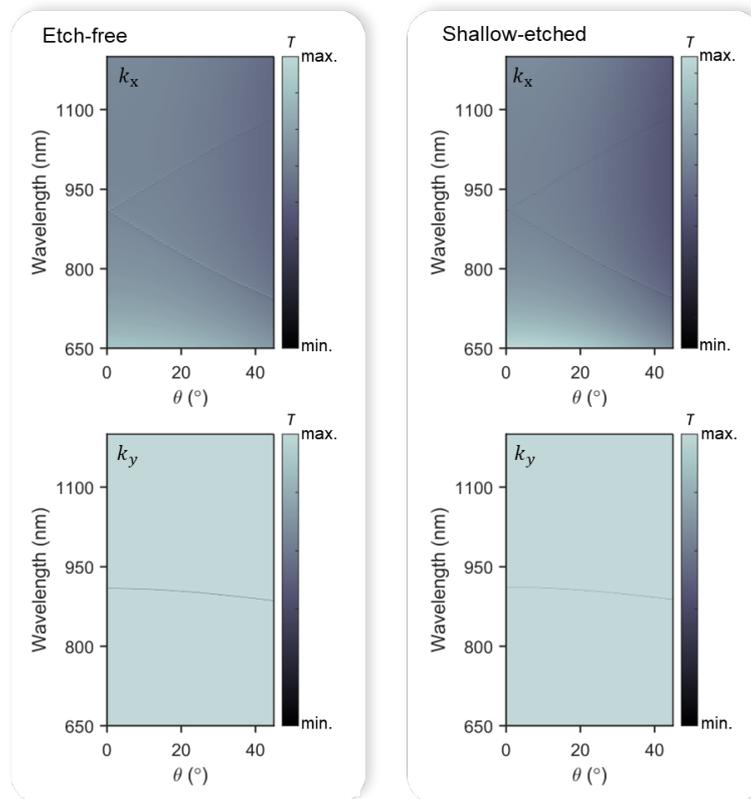

Figure S4. Comparison between the dispersion of transmission spectra of etch-free and shallow-etched structures shown in Figure. 1b.

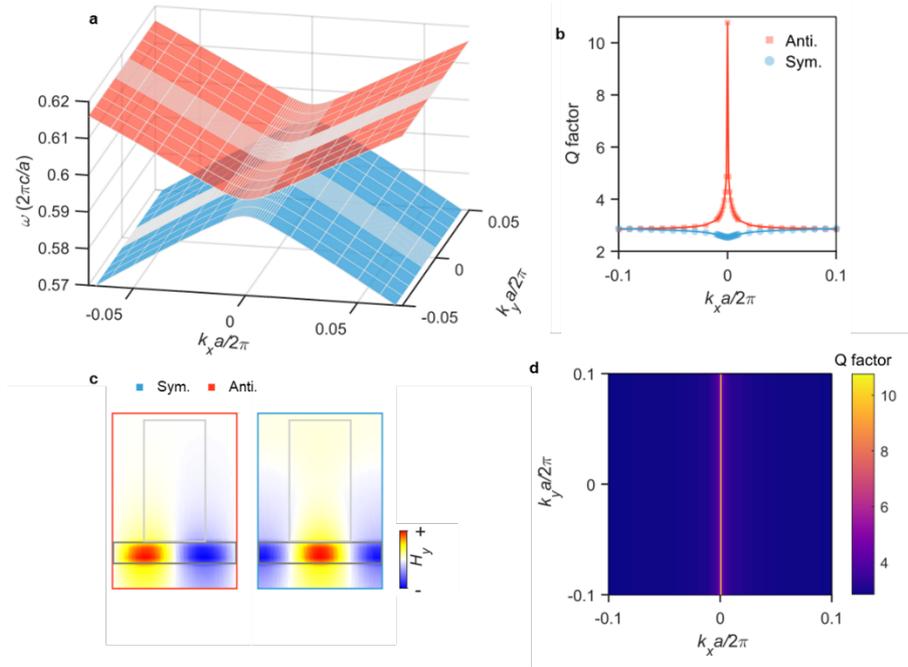

Figure S5. Calculated optical properties of TM mode of etch-free vdW grating, corresponding to the results of TE mode which are shown in Fig1. **c-e**.

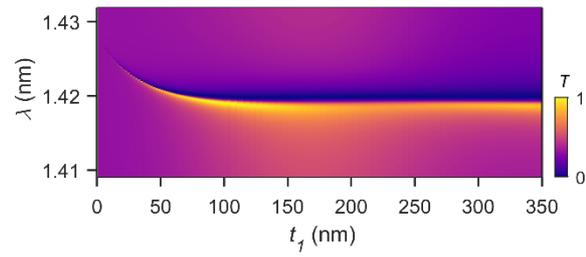

Figure S6. The evolution of transmission spectra (TE) with the thickness of ZEP thickness.

**Fitting results**

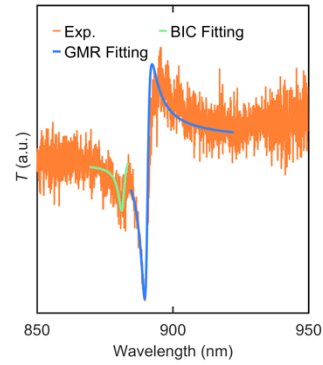

Figure S7. Fitting results for the split BIC and GMR TE modes (results in Figure **1f** and Figure **4b** in the main text) with at 1-degree inclined incidence angle. GMR and (quasi)-BIC modes share the similar linewidth.

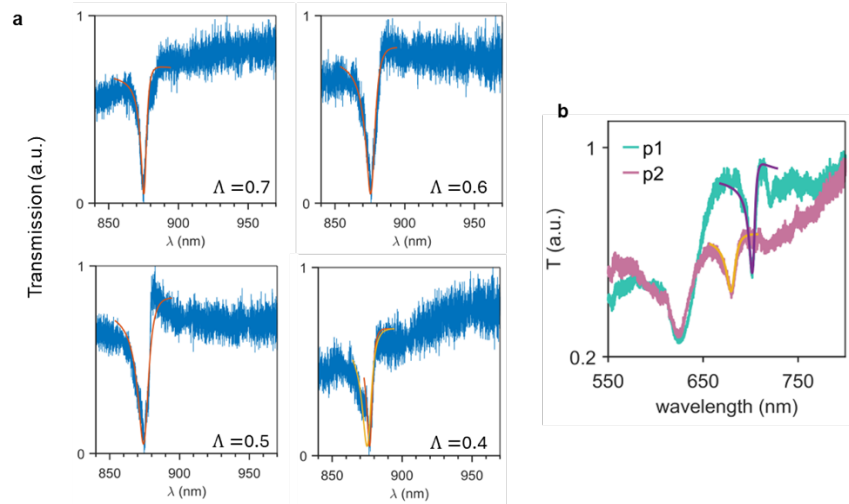

Figure S8. Fitting the transmission spectra with the Fano formula $F(\omega) = R_0 + \frac{A(q+(\frac{\omega-\omega_0}{\gamma})^2)}{(1+(\frac{\omega-\omega_0}{\gamma})^2)}$. q is the Fano factor, and $\omega_0$ ($\gamma$) is the resonant frequency (linewidth) of the mode. (**a**) Fitting results corresponding to Fig. **2l** in the main text. (**b**) Fitting results corresponding to Fig. **2i** in the main text.

| Λ | q | $\omega_0$ | Γ ($\gamma = \Gamma/2$) | $R_0$ | A |
|---|---|---|---|---|---|
| 0.7 | -0.2 | 1.416 | 0.008 | 0.05 | 0.65 |
| 0.6 | -0.2 | 1.415 | 0.013 | 0.05 | 0.75 |
| 0.5 | -0.2 | 1.417 | 0.015 | 0.05 | 0.75 |
| 0.4 | -0.2 | 1.4135 & 1.416 | 0.008 % 0.13 | 0.05 | 0.6 |

Table S1. Fitting parameters for Figure **S5a**.

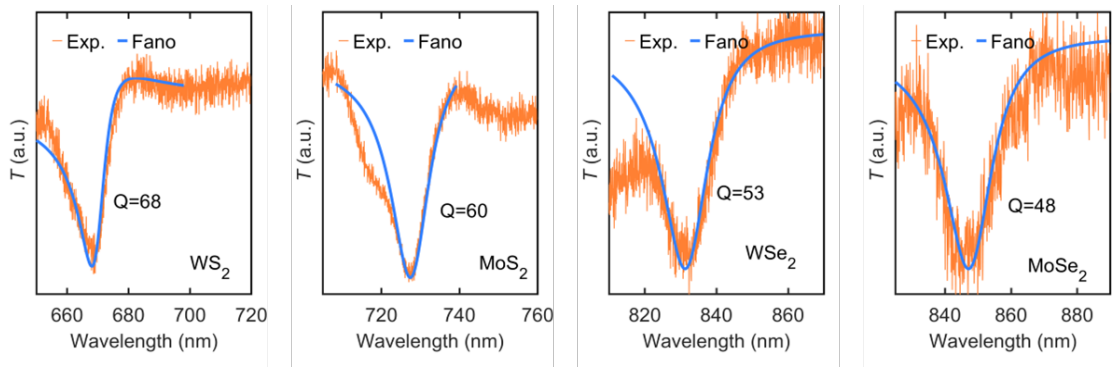

Figure S9. Fano fitting of transmission TM modes (normal) for four samples shown in Figure **3** in the main text.

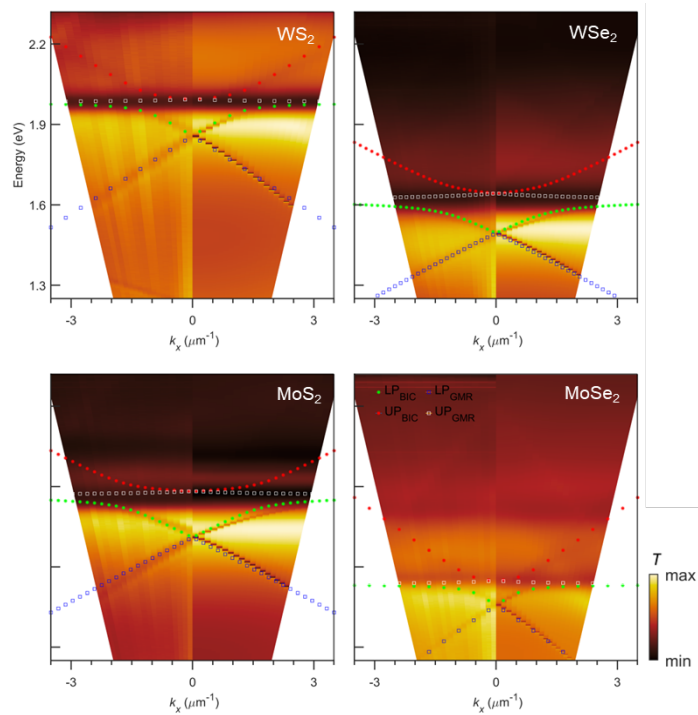

Figure S10. Polaritonic dispersions for the results in Figure **3** in the main text. The green and red scatters correspond to the hybrid LP BIC and UP BIC branches while the blue and white scatters correspond to the hybrid LP GMR and UP GMR branches.

**Experimental fabrication**

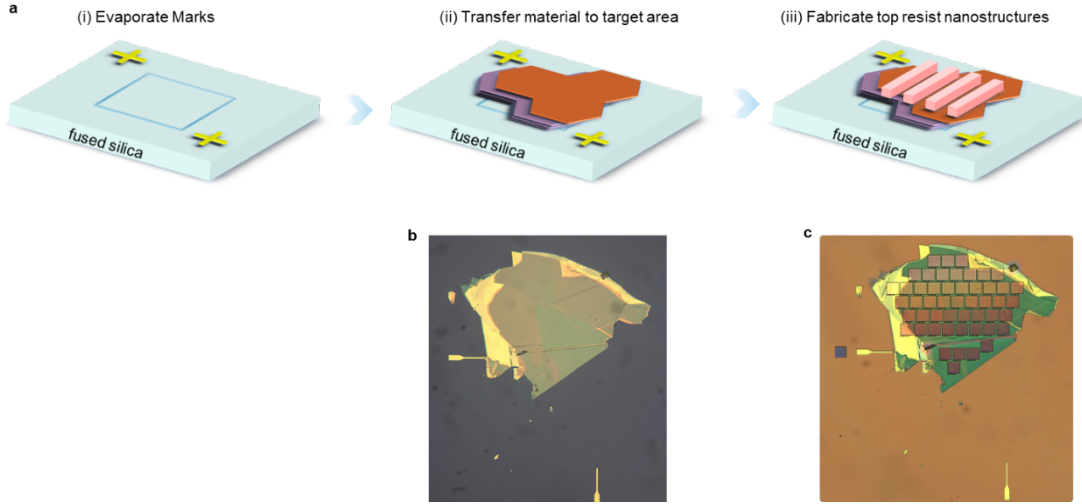

Figure S11. (a) Fabrication procedure of etch-free vdW metasurfaces. (b) Microscopic images of transferred vdW materials on targeted area marked by Au arrows. (c) Microscopic image of the fabricated structures after the EBL and developing processes.

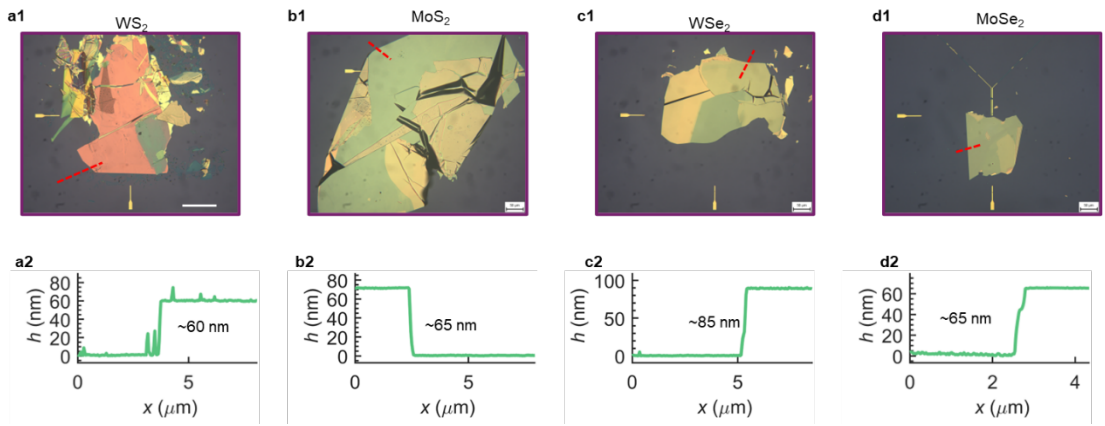

Figure S12. The thickness of four TMDC materials. (**a1-d1**) Microscopic of the bulk $WS_2$, $MoS_2$, $WSe_2$, and $MoSe_2$. (**a2-d2**) Corresponding measured thickness from the above four materials with the measurement region marked by the red lines in (**a1-d1**).

**Homebuilt quasi-collimated transmission measurements system**

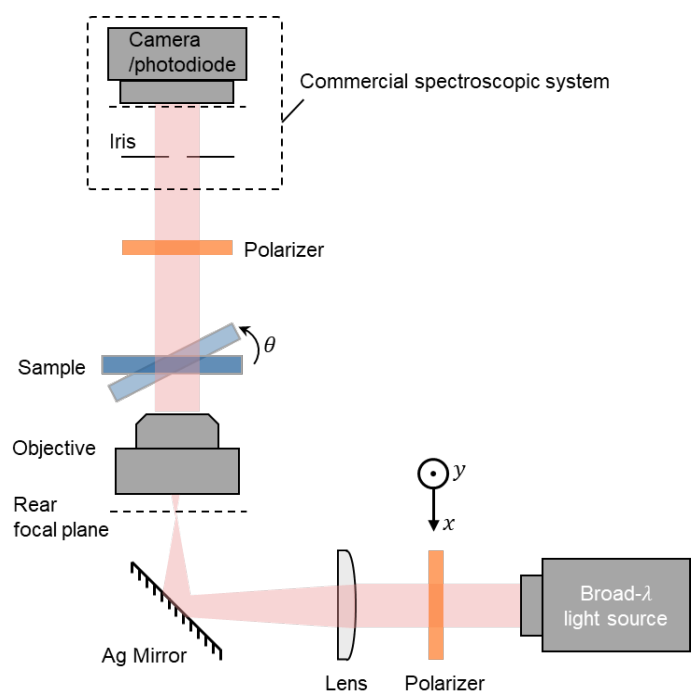

Figure S12. Home-built setup for the transmission measurements. The broad-wavelength light first passes through a linear polarizer, Plano-convex lens (20X), the Ag mirror in sequence, to be focused on the rear focal plane of a 5X objective. Then the collimated light would pass through the sample for the measurements. Another linear polarizer would filter different polarization of the transmitted light. Finally, the light would be measured in the commercial spectroscopic system (HORIBA LabRAM HR Evolution system).

**Active devices**

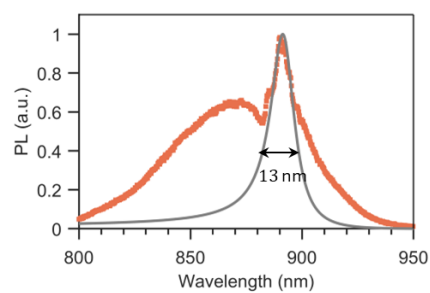

Figure S13. Fitting enhanced indirect bandgap emission spectrum for Fig. **4e** in main text.

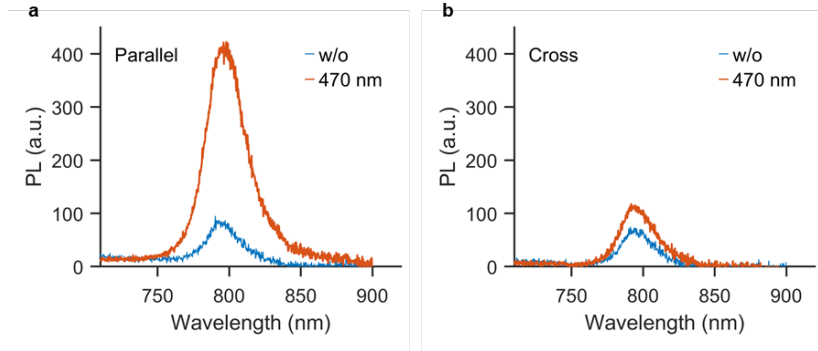

Figure S14. PL emission of heterostructure shown in Fig. **5** in main text. Red lines: with etch-free structure. Blue lines: without etch-free structure.

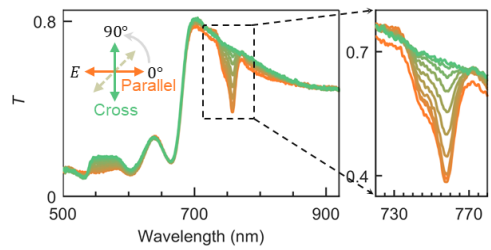

Figure S14. Polarization-dependent transmission spectra of the etch-free MoS$_2$ metasurface.